\documentclass[preprints,article,accept,pdftex,moreauthors]{Definitions/mdpi} 
\usepackage{tcolorbox}

\firstpage{1} 
\pubvolume{1}
\issuenum{1}
\articlenumber{0}
\pubyear{2025}
\copyrightyear{2025}
\datereceived{ } 
\daterevised{ } 
\dateaccepted{ } 
\datepublished{ } 
\hreflink{https://doi.org/} 

\Title{Kinematic Anisotropies in PTA Observations: Analytical Toolkit}

\TitleCitation{Title}

\newcommand{\Eqref}[1]{Eq.~\eqref{#1}}

\Author{Maximilian Blümke$^{1,*}$\,\orcidB{}, Kai Schmitz$^{1,2,*}$\,\orcidD{}, Tobias Schröder$^{1,3}$\,\orcidE{}, Deepali Agarwal$^{4}$\,\orcidA{} and  Joseph D. Romano$^{4}$\,\orcidC{}}

\AuthorNames{Firstname Lastname, Firstname Lastname and Firstname Lastname}

\isAPAStyle{%
       \AuthorCitation{Lastname, F., Lastname, F., \& Lastname, F.}
         }{%
        \isChicagoStyle{%
        \AuthorCitation{Lastname, Firstname, Firstname Lastname, and Firstname Lastname.}
        }{
        \AuthorCitation{Lastname, F.; Lastname, F.; Lastname, F.}
        }
}

\address{%
$^{1}$ \quad Institute for Theoretical Physics, University of Münster,
 Wilhelm-Klemm-Straße 9, 48149 Münster, Germany\\
$^{2}$ \quad Kavli IPMU (WPI), UTIAS, The University of Tokyo, 5-1-5 Kashiwanoha, Kashiwa, Chiba 277-8583, Japan\\
$^{3}$ \quad Department of Physics, McGill University, Montreal, QC, H3A 2T8, Canada\\
$^{4}$ \quad Department of Physics and Astronomy, University of Texas Rio Grande Valley, One West University Boulevard, Brownsville, TX 78520, USA
}

\corres{Correspondence: maximilian.bluemke@uni-muenster.de, kai.schmitz@uni-muenster.de}

\abstract{The reported evidence for an isotropic gravitational-wave background (GWB) from pulsar timing array (PTA) collaborations has motivated searches for extrinsic and intrinsic anisotropies. Kinematic anisotropies may arise as a consequence of a boosted observer moving with respect to the frame in which the GWB appears isotropic. In this work, we present an analytical toolbox to describe the effects of kinematic anisotropies on the overlap reduction function. Our analytical results differ from previous findings at the quadrupole order and are detailed in three appendices. For the first time, we also derive the corresponding auto-correlation using two approaches, taking the pulsar distances to be infinite or finite, respectively. Our formulas can be used in forecasts or Bayesian analysis pipelines.}


\begin{document}


\section{Introduction}

Recent results from several pulsar timing array (PTA) collaborations contain weak to strong evidence for an isotropic gravitational-wave background (GWB) signal~\cite{Reardon:2023gzh,EPTA:2023fyk,Miles:2024seg,NANOGrav:2023gor,Xu:2023wog}. The measured frequency spectrum and inter-pulsar correlation pattern are found to be consistent with both supermassive black-hole binary populations~\cite{NANOGrav:2023hfp} and certain cosmological models (such as, e.g., cosmic inflation, a cosmological first-order phase transition, or cosmic strings)~\cite{NANOGrav:2023hvm,EPTA:2023xxk}, but current data cannot yet distinguish between them. Forthcoming combined analyses from the International Pulsar Timing Array are expected to improve sensitivity and may provide more insights into the signal's origin~\cite{Antoniadis:2022pcn,InternationalPulsarTimingArray:2023mzf}.

A promising avenue to distinguish the origin of the signal lies in the search for anisotropies in the GWB. Supermassive black-hole binaries are expected to generate anisotropies at the 1--20$\%$ level at low multipoles~\cite{Taylor:2013esa,Mingarelli:2013dsa,Mingarelli:2017fbe,Semenzato:2024mtn,Sah:2024oyg,Sah:2024etc,Allen:2024mtn}. These anisotropies arise from the inhomogeneous distribution of matter and large-scale structure, as well as from shot noise associated with the discrete nature of the contributing sources. Recent PTA observations, however, have not yet revealed any significant level of anisotropy~\cite{NANOGrav:2023tcn,Grunthal:2024sor}. Anisotropies in the cosmological GWB are expected to arise from two main sources: anisotropies at source production and propagation effects, including the Sachs--Wolfe and integrated Sachs--Wolfe effects~\cite{Contaldi:2016koz,Bartolo:2019yeu,Cusin:2018avf,Pitrou:2019rjz}. Such anisotropies are predicted for signals produced in the early Universe, including those originating from inflation~\cite{Bartolo:2019oiq,Bartolo:2019yeu,Adshead:2020bji,Malhotra:2020ket,Dimastrogiovanni:2021mfs}, phase transitions~\cite{Geller:2018mwu,Kumar:2021ffi,Li:2021iva}, topological defects~\cite{Cai:2021dgx,Liu:2020mru,Jenkins:2018nty,Kuroyanagi:2016ugi}, or primordial black holes~\cite{Bartolo:2019zvb}. Due to the extremely weak interactions of gravitons with matter, the properties of the primordial GWB depend sensitively on the underlying production mechanism, though anisotropies from early-Universe phenomena are expected to be small with typical amplitudes of around $C_{\ell}\sim 10^{-5}$ (see~\cite{LISACosmologyWorkingGroup:2022kbp} for a review on GWB sources and their anisotropies and review article~\cite{Sesana:2025udx} for GWB sources in the nHz frequency range).

In addition, our peculiar motion with respect to the cosmological rest frame---established through CMB~\cite{Planck:2018nkj} as well as quasar and radio galaxy observations~\cite{Secrest:2020has,Dalang:2021ruy,Secrest:2022uvx}---induces Doppler-boosted kinematic anisotropies and corrections to the frequency spectrum of the monopole signal~\cite{Allen:1996gp,Cusin:2022cbb,Tasinato:2023zcg,Cruz:2024svc,Mentasti:2025ywl}. Kinematic anisotropy---an extrinsic, large-scale dipolar anisotropy---arises from the Doppler modulation of the GWB intensity, producing an enhancement in the direction of the observer's motion and a corresponding suppression in the opposite direction. The motion of the Solar System barycenter toward the so-called ``Great Attractor" was first inferred from CMB observations, and the latest results are a best-fit amplitude of 369.82 km s$^{-1}$ in the direction $\ell=264^{\circ}, b=48^{\circ}$ in galactic coordinates~\cite{Planck:2018nkj}. Later, it was confirmed using astrophysical tracers such as radio galaxies and quasars. A notable dipole tension exists: while the directions inferred from these probes agree, their amplitudes are inconsistent at the 5$\sigma$ level~\cite{Secrest:2022uvx,Wagenveld:2025ewl,Secrest:2025wyu}. 

Since the GWB serves as both an astrophysical and cosmological probe, the same kinematic anisotropy should imprint itself on gravitational-wave measurements. If the GWB should originate from supermassive black-hole binaries, the kinematic dipole may easily be superseded by an intrinsic dipole \cite{Taylor:2020zpk,Becsy:2022pnr}. On the other hand, if it should be of cosmological origin, the kinematic dipole may represent the leading contribution, similar to the case of the CMB \cite{Bartolo:2019yeu,LISACosmologyWorkingGroup:2022kbp,Cruz:2024svc}. This expectation nourishes the hope that the confirmation of kinematic anisotropies, alongside a null measurement of intrinsic anisotropies, could ultimately provide evidence in favor of the cosmological interpretation of the signal. Motivated by these prospects, several ongoing efforts aim at modelling the kinematic effects in the GWB~\cite{Cusin:2022cbb,ValbusaDallArmi:2022htu,Mentasti:2025ywl} and at developing methods to disentangle intrinsic from extrinsic anisotropies~\cite{ValbusaDallArmi:2022htu}. The first search for the kinematic GWB dipole was conducted with ground-based GW interferometers, and did not reveal any significant evidence of anisotropy~\cite{Chung:2022xhv}.

As we will discuss in this article, kinematic anisotropies represent in particular a well-defined (albeit challenging) observational target for PTA observations: the orientation of the leading dipole contribution is fixed by the direction of motion of the observer; its amplitude is determined by the velocity of the observer; and both quantities (i.e., the direction and the strength of the dipole) ought to be consistent with CMB measurements. Furthermore, kinematic anisotropies are characterized by fixed relations between the dipole and all higher-multipole contributions. A confirmation of (a subset of) these properties could support the cosmological interpretation of the signal.

PTA searches for an isotropic GWB rely on the Hellings--Downs curve~\cite{Hellings:1983fr}, i.e., the cross-correlation between the timing residuals of pairs of pulsars as a function of their angular separation. However, deviations from the Hellings--Downs curve can occur in PTA measurements due to the stochastic nature of the GWB, a phenomenon known as cosmic variance~\cite{Allen:2022dzg,Bernardo:2023bqx,Agarwal:2024hlj,Domcke:2025esw}. In addition, PTA observations search for GWB anisotropies, taking advantage of distinctive deviations from the Hellings--Downs curve~\cite{Mingarelli:2013dsa,Taylor:2013esa,Ali-Haimoud:2020ozu,alihaimoud2021}. In particular, the effect of dipole anisotropy has been analyzed in~\cite{Anholm:2008wy,Mingarelli:2013dsa}. The impact of kinematic GWB anisotropy (dipole and quadrupole) on the cross-correlation of PTA timing residuals was first computed by~\citet{Tasinato:2023zcg}. These results were subsequently applied to the 15-year dataset from the NANOGrav collaboration~\cite{Cruz:2024svc}. The derived constraints on the GWB amplitude and spectral shape were found to be consistent with those reported in Ref.~\cite{NANOGrav:2023gor}, and a Bayesian upper limit at the $95\%$ credible level on the velocity of the observer, $v$, in units of the speed of light, $c$, was established as $\beta = v/c < 0.297$, assuming a cosmological origin of the GWB.

Here, we present a revised expression for the effect of the quadrupole on the cross-correlation of timing residuals that differs from earlier results. Additionally, we account for the impact of kinematic anisotropy on auto-correlations~\cite{Cordes:2024oem, Anholm:2008wy}. While the pulsar terms are often neglected in GWB cross-correlation analyses, they play an important role in auto-correlations and must be properly included. This is already true in the standard case, $\beta =0$, where the pulsar terms yield  an additional factor of $2$ in the auto-correlation coefficient. Here, we generalize the discussion to nonzero relative velocity. Our analytical framework for kinematic effects on the correlation curve provides a useful toolkit for PTA searches for kinematic anisotropies as well as for probing both the nature of the GWB source and the motion of the observer.

The rest of the article is structured as follows. After a brief introduction to the kinematic effects on the cosmological GWB, we introduce the overlap reduction function (ORF) in Sec.~\ref{sec:SecII}, which acts as a geometrical factor accounting for correlations among pulsar pairs. Section~\ref{sec:SecIIa} presents the expressions for the kinematic contribution to the cross-correlation ORF, while the auto-correlation ORF is discussed in Sec.~\ref{sec:SecIIb}. 
In Sec.~\ref{sec:SecIIc}, we apply the results of the previous section to compute the change in the shape of isotropic ORF, using the array of pulsars from the NANOGrav 15-year analysis~\cite{NANOGrav:2023gor}.
We conclude in Sec.~\ref{sec:SecIII} with a summary and outlook. Several appendices provide detailed analytical calculations underlying the results summarized in the main text. We use the Einstein summation convention for repeated spatial indices $i$, $j$, $\cdots$, and work with units where the speed of light is set to unity, $c=1$.

\section{Effect of kinematic anisotropy on the gravitational-wave background}\label{sec:SecII}

We begin with the derivation of the ORF, including the effect of the relative motion between the observer's rest frame $\mathcal{S}$ and the rest frame of the GWB source $\mathcal{S}'$. We assume the GWB to be homogeneous and isotropic in the latter frame.
Our treatment closely follows Refs.~\cite{Cusin:2022cbb, Tasinato:2023zcg, Cruz:2024svc}. 

To characterize the GWB spectrum,  we start with the GW density parameter in the frame $\mathcal{S}'$:
\begin{align}
\Omega_{\rm GW}'(f)  = \frac{4\pi^2 f^3}{3H_0^2} I'(f) \, ,
\end{align}
where $I'(f)$ denotes the GW intensity. Because the GWB is assumed to be isotropic in this frame, $I'(f)$ depends only on frequency, but not on the direction $\hat{\boldsymbol{n}}$ on the celestial sphere. Let us now boost from the isotropic frame $\mathcal{S}'$ to the observer frame $\mathcal{S}$, which moves with a velocity of $\boldsymbol{v} = \beta \hat{\boldsymbol{v}}$ relative to $\mathcal{S}'$. A frequency $f'$ in the frame $\mathcal{S}'$
is observed in the frame $\mathcal{S}$ as the Doppler-shifted frequency
\begin{align}
    f(\hat{\boldsymbol{n}}) = \mathcal{D}(\hat{\boldsymbol{n}}) f' && \text{where} && \mathcal{D}(\hat{\boldsymbol{n}}) = \frac{\sqrt{1- \beta^2}}{1 - \beta\, \hat{\boldsymbol{n}}\cdot \hat{\boldsymbol{v}}} \, .
\end{align}
Using the independence of the number of gravitons on the chosen frame, the authors of Ref.~\cite{Cusin:2022cbb} showed that the density parameters in the two different frames are related via
\begin{align}
    \Omega_{\rm GW}(f, \hat{\boldsymbol{n}})= \mathcal{D}^4(\hat{\boldsymbol{n}})\, \Omega'_{\rm GW}(\mathcal{D}^{-1}(\hat{\boldsymbol{n}})f) \,.
\end{align}
For the GW intensity, this implies
\begin{align}
    \frac{I(f,\hat{\boldsymbol{n}})}{I'(f)} = \frac{\mathcal{D}(\hat{\boldsymbol{n}})I'(\mathcal{D}^{-1}(\hat{\boldsymbol{n}})f)}{I'(f)} \, .
\end{align}
Assuming that the relative velocity is small, $\beta \ll 1$, as expected based on the value inferred from CMB measurements, $\beta_{\rm CMB} \sim 10^{-3}$, we expand the expression in powers of $\beta$. Following Ref.~\cite{Tasinato:2023zcg}, this yields
\begin{equation}
\begin{aligned}
    \label{eq:taylorExpansionInBeta}
    \frac{I(f, \hat{\boldsymbol{n}})}{ I' (f)}=& \left[ 1 - \frac{\beta^2}{6} (1 - n_I^2 - \alpha_I) \right] + \beta (1- n_I) \left( \hat{\boldsymbol{n}} \cdot \hat{\boldsymbol{v}} \right) \\
    &\hspace{0.5in}+ \frac{\beta^2}{2} (2- 3 n_I + n_I^2 + \alpha_I)\left( (\hat{\boldsymbol{n}}\cdot \hat{\boldsymbol{v}})^2 - \frac{1}{3} \right)  + \mathcal{O}(\beta^3) \, .
\end{aligned}
\end{equation}
Here, $n_I$ and $\alpha_I$ denote the frequency-dependent spectral index and its logarithmic derivative:
\begin{align}\label{eq:spectralParameter}
    n_I(f) = \frac{d \ln I'}{d \ln f} && \text{and} && \alpha_I(f) = \frac{d n_I}{d \ln f} \, .
\end{align}
It should also be mentioned that the Taylor expansion in \Eqref{eq:taylorExpansionInBeta} is not valid for strong scale dependence as recently shown in Ref.~\cite{Mentasti:2025ywl}.
Let us now compute the corresponding ORF \cite{Tasinato:2023zcg,Cruz:2024svc}:
\begin{align}\label{eq:fullORF}
   \Gamma_{ab} = \frac{3}{8 \pi}\int d^2 \hat{\boldsymbol{n}} \, \sum_\lambda D_a^\lambda (\hat{\boldsymbol{n}}) D_b^{\lambda} (\hat{\boldsymbol{n}}) \left[ 1- e^{2\pi i  f {L_a} (1+\hat{\boldsymbol{n}}\cdot \hat{\boldsymbol{x}}_a)}\right] \left[ 1- e^{-2\pi i f {L_b} (1+\hat{\boldsymbol{n}}\cdot \hat{\boldsymbol{x}}_b)} \right] \frac{I(f, \hat{\boldsymbol{n}})}{I'(f)} \,,
\end{align}
where $L_a$ and $L_b$ denote the distances of pulsars $a$ and $b$ from the solar system and where
\begin{equation}
D^\lambda_a(\hat{\boldsymbol{n}})
\equiv \frac{1}{2}\frac{\hat x_a^i \hat x_b^j e_{ij}^\lambda(\hat{\boldsymbol{n}})}
{1 + \hat{\boldsymbol{x}}_a\cdot \hat{\boldsymbol{n}}}\,,
\qquad \lambda\equiv\{+,\times\}\,,
\end{equation}
are the antenna pattern functions for pulsar $a$  with $e^\lambda_{ij}(\hat{\boldsymbol{n}})$ denoting the GW polarization tensors,
see \Eqref{e:polarization_tensors}.
For realistic pulsar distances and the frequency range of interest, we have $fL \gg 1$. Consequently, the rapidly oscillating terms in the integral can be neglected, reducing the expression to
\begin{align} \label{eq:fullORF3}
     \Gamma_{ab} =& \frac{3}{8 \pi}(1+\delta_{ab})\int d^2 \hat{\boldsymbol{n}} \, \sum_\lambda D_a^\lambda (\hat{\boldsymbol{n}}) D_b^{\lambda} (\hat{\boldsymbol{n}}) \frac{I(f, \hat{\boldsymbol{n}})}{I'(f)} \, .
\end{align}

\subsection{Cross-correlations}\label{sec:SecIIa}

Let us first consider cross-correlations, i.e., $a\neq b$. Inserting the $\beta$-expansion of \Eqref{eq:taylorExpansionInBeta}, the ORF takes the form \cite{Cruz:2024svc}
\begin{align}\label{eq:fullORF2}
   \Gamma_{ab} =\left[ 1 - \frac{\beta^2}{6} (1 - n_I^2 - \alpha_I) \right] \Gamma_{ab}^{(0)} + \beta (1- n_I)  \Gamma_{ab}^{(1)} + \frac{\beta^2}{2} (2- 3 n_I + n_I^2 + \alpha_I) \Gamma_{ab}^{(2)}  + \mathcal{O}(\beta^3) \, ,
\end{align}
where we introduced three contributions, referred to as the monopole, dipole, and quadrupole contributions in the following:
\begin{align}
    \Gamma_{ab}^{(0)} & = \frac{3}{8 \pi} \int d^2\hat{\boldsymbol{n}} \, \sum_\lambda D_a^\lambda(\hat{\boldsymbol{n}})D_b^\lambda(\hat{\boldsymbol{n}})\,, \label{eq:Monopoledef}\\
    \Gamma_{ab}^{(1)} & = \frac{3}{8 \pi} \int d^2\hat{\boldsymbol{n}} \, (\hat{\boldsymbol{n}}\cdot \hat{\boldsymbol{v}})\sum_\lambda D_a^\lambda(\hat{\boldsymbol{n}}) D_b^\lambda(\hat{\boldsymbol{n}}) \,, \label{eq:Dipoledef}\\
    \Gamma_{ab}^{(2)} & = \frac{3}{8 \pi} \int d^2\hat{\boldsymbol{n}} \, \left[(\hat{\boldsymbol{n}} \cdot\hat{\boldsymbol{v}})^2 - \frac{1}{3} \right]\sum_\lambda D_a^\lambda(\hat{\boldsymbol{n}}) D_b^\lambda(\hat{\boldsymbol{n}}) \label{eq:Quadrupoledef}\,.
\end{align}
For $\beta = 0$, we have $I(f, \hat{\boldsymbol{n}})/{I'(f)}=1$. The dipole and quadrupole vanish and the ORF reduces to the well-known Hellings--Downs correlation (see appendix \ref{appendix:monopole} for a detailed derivation):
\begin{align}
\label{eq:GammaHD}
    \Gamma_{ab}^{(0)} = \frac{1}{2} - \frac{1}{4} x_{ab} + \frac{3}{2} x_{ab} \ln \left( x_{ab} \right)\, ,
\end{align}
where 
\begin{align}
    x_{ab} \equiv \frac{1 - \cos \xi}{2} && \text{and} && \cos(\xi)=\hat{\boldsymbol{x}}_a \cdot\hat{\boldsymbol{x}}_b \,.
\end{align}
For $\beta>0$, \Eqref{eq:fullORF2} shows that the monopole receives corrections of $\mathcal{O}(\beta^2)$, the dipole appears at $\mathcal{O}(\beta)$, and the quadrupole at $\mathcal{O}(\beta^2)$. The dipole contribution was first derived in Ref.~\cite{Anholm:2008wy} and reads:
\begin{align}
    \Gamma_{ab}^{(1)} & = \left( -\frac{1}{8} - \frac{3}{4}x_{ab} - \frac{3}{4}\frac{x_{ab}}{\left( 1- x_{ab}\right)}\ln x_{ab}\right)   \left[ \hat{\boldsymbol{v}} \cdot\hat{\boldsymbol{x}}_a + \hat{\boldsymbol{v}} \cdot\hat{\boldsymbol{x}}_b \right]. \label{eq:dipoleHD}
\end{align}
A detailed derivation of this result is provided in appendix \ref{appendix:dipole}.

Let us now turn to the quadrupole contribution to the ORF. Following the computation detailed in appendix \ref{appendix:quadrupole}, one finds
\begin{tcolorbox}
\vspace{-0.5cm}
\begin{equation}\label{eq:ourGamma2ResultimText}
\begin{split}
\Gamma_{ab}^{(2)} 
&= \frac{1}{\left( x_{ab} -1 \right)^2} \bigg\{
    \left( 
        \frac{1}{4}x_{ab}^3 - x_{ab}^2 + \frac{3}{4} x_{ab} \ln x_{ab} 
        + \frac{1}{2} x_{ab} + \frac{1}{4} 
    \right) 
    \left[ 
        (\hat{\boldsymbol{v}}\cdot\hat{\boldsymbol{x}}_a)^2 
        + (\hat{\boldsymbol{v}}\cdot\hat{\boldsymbol{x}}_b)^2
    \right] \\
&\hspace{0.5in} 
    + \left( 
        \frac{3}{2} x_{ab}^2 \ln x_{ab} - \frac{39}{20} x_{ab}^2 
        + \frac{12}{5} x_{ab} - \frac{9}{20}
    \right) 
    \left[
        (\hat{\boldsymbol{v}}\cdot\hat{\boldsymbol{x}}_a)
        (\hat{\boldsymbol{v}}\cdot\hat{\boldsymbol{x}}_b)
    \right] \\
&\hspace{0.5in}
    + \left(
        x_{ab}^3 \ln x_{ab} - \frac{22}{15} x_{ab}^3 
        - \frac{1}{2} x_{ab}^2 \ln x_{ab} + \frac{35}{12} x_{ab}^2 
        - \frac{1}{2} x_{ab} \ln x_{ab} - \frac{43}{30} x_{ab} 
        - \frac{1}{60}
    \right)
\bigg\}\,.
\end{split}
\end{equation}
\end{tcolorbox}
\noindent
This result disagrees with the expression obtained in Refs.~\cite{Tasinato:2023zcg, Cruz:2024svc}, which we reproduce here for completeness (with the normalization adapted to our conventions):
\begin{equation}\label{eq:tasinatoGamma2Result}
\begin{split}
\tilde{\Gamma}_{ab}^{(2)}
&= \frac{3}{4}\left( 
        \frac{ 13 x_{ab}-3}{10 (1-x_{ab})} 
        + \frac{x_{ab}^2 \ln x_{ab}}{(1 - x_{ab})^2} 
    \right) 
    \left[ 
        (\hat{\boldsymbol{v}}\cdot\hat{\boldsymbol{x}}_a) 
        (\hat{\boldsymbol{v}}\cdot\hat{\boldsymbol{x}}_b) 
    \right] \\
&\hspace{0.5in}
    + \left( 
        \frac{1 + 2x_{ab} - 4 x_{ab}^2 + x_{ab}^3 + 3 x_{ab} \ln x_{ab}}
        {8 (1 - x_{ab})^2} 
    \right) 
    \left[
        (\hat{\boldsymbol{v}}\cdot \hat{\boldsymbol{x}}_a)^2 
        + (\hat{\boldsymbol{v}}\cdot \hat{\boldsymbol{x}}_b)^2 
    \right] \,.
\end{split}
\end{equation}

\begin{figure}
    \isPreprints{\centering}{} 
    \includegraphics[width=0.75\linewidth]{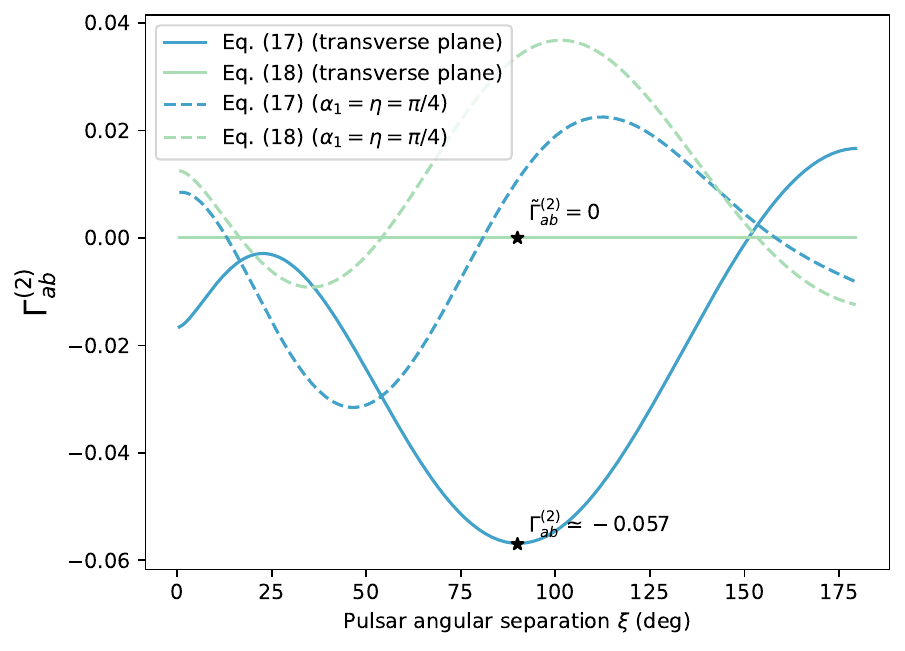}
    \caption{Comparison of the solutions for the quadrupole contribution from Ref.~\cite{Cruz:2024svc} and our analysis (adjusting for normalization). The solid lines indicate a system in which the pulsar direction vectors lie in the plane transverse to the velocity vector, while the dashed plots are for the case of $\alpha_1 = \eta = \pi /4$ (angles for $\hat{\boldsymbol{v}}$; see \Eqref{eq:vinspherical}). The black stars indicate the result considered for the numerical solution in \Eqref{eq:smallNumericalInOrthogonal} and the vanishing solution for the same configuration in \Eqref{eq:tasinatoGamma2Result}.}
    \label{fig:ComparisonQuadrupole}
\end{figure}

In Fig.~\ref{fig:ComparisonQuadrupole}, we compare our result $\Gamma_{ab}^{(2)}$ from \Eqref{eq:ourGamma2ResultimText} (blue curves) with the result $\tilde{\Gamma}_{ab}^{(2)}$ from \Eqref{eq:tasinatoGamma2Result} (teal curves) for two benchmark geometries.
The first benchmark geometry consists of a configuration in which the observer velocity $\hat{\boldsymbol{v}}$ points along the $y$-axis, while the pulsars lie in the $xz$-plane, i.e., in the plane orthogonal to the velocity direction. This ``transverse-plane" geometry corresponds to the solid curves shown in Fig.~\ref{fig:ComparisonQuadrupole}. In this case, the integral defining the quadrupole,
\begin{align} \label{eq:integralGamma2}
     \Gamma_{ab}^{(2)} = \frac{3}{8 \pi} \int d^2 \hat{\boldsymbol{n}} \,\left[(\hat{\boldsymbol{n}}\cdot\hat{\boldsymbol{v}})^2 - \frac{1}{3}\right] \sum_\lambda D_a^\lambda(\hat{\boldsymbol{n}}) D_b^\lambda(\hat{\boldsymbol{n}}) \, ,
\end{align}
reduces to the simpler form
\begin{equation}
\begin{split}
\Gamma_{ab}^{(2)}(\xi) 
&= \frac{3}{16 \pi} 
    \int_{-1}^{1} dx \, 
    \int_{0}^{2 \pi} d\phi \, 
    \left[(1-x^2) \sin^2 \phi - \frac{1}{3}\right] 
    \frac{1}{2} (1-x) \\
&\hspace{0.5in}
    \times \left[
        (1 - x\cos \xi 
         - \sqrt{1 - x^2}\, \sin \xi \cos \phi)
        - \frac{2 \sin^2 \xi \sin^2 \phi}
        {1 + x \cos \xi + \sqrt{1 - x^2} \cos \phi}
    \right]\, ,
\end{split}
\end{equation}
where $\xi$ is the angle between the pulsars, and $x = \cos\theta$, with $(\theta,\phi)$ being the spherical coordinates associated with the direction $\hat{\boldsymbol{n}}$. Evaluating this expression numerically for $\xi = \pi/2$, we obtain
\begin{align}\label{eq:smallNumericalInOrthogonal}
    \Gamma_{ab}^{(2)} \simeq -0.057 \, .
\end{align}
While this matches our result in \Eqref{eq:ourGamma2ResultimText}, it does not agree with \Eqref{eq:tasinatoGamma2Result}. The latter expression vanishes identically, since it contains only terms proportional to $\hat{\boldsymbol{v}}\cdot\hat{\boldsymbol{x}}_a$ and $\hat{\boldsymbol{v}}\cdot\hat{\boldsymbol{x}}_b$, both of which are zero in the configuration where $\hat{\boldsymbol{x}}_{a,b} \perp \hat{\boldsymbol{v}}$. This is evident from Fig.~\ref{fig:ComparisonQuadrupole}, where $\tilde{\Gamma}_{ab}^{(2)}$ vanishes for all values of $\xi$, whereas our expression $\Gamma_{ab}^{(2)}$ vanishes only at a single angular separation. Contributions to the quadrupole that are non-vanishing for this configuration are viable as they can arise due to the $-1/3$ term in \Eqref{eq:integralGamma2} as well as from terms proportional to $(\hat{\boldsymbol{v}}\cdot\hat{\boldsymbol{v}}) 
(\hat{\boldsymbol{x}}_a\cdot\hat{\boldsymbol{x}}_a)$ or $(\hat{\boldsymbol{v}}\cdot\hat{\boldsymbol{v}}) (\hat{\boldsymbol{x}}_b\cdot\hat{\boldsymbol{x}}_b)$.  

\begin{figure} 
    \isPreprints{\centering}{} 
    \includegraphics[scale=0.5]{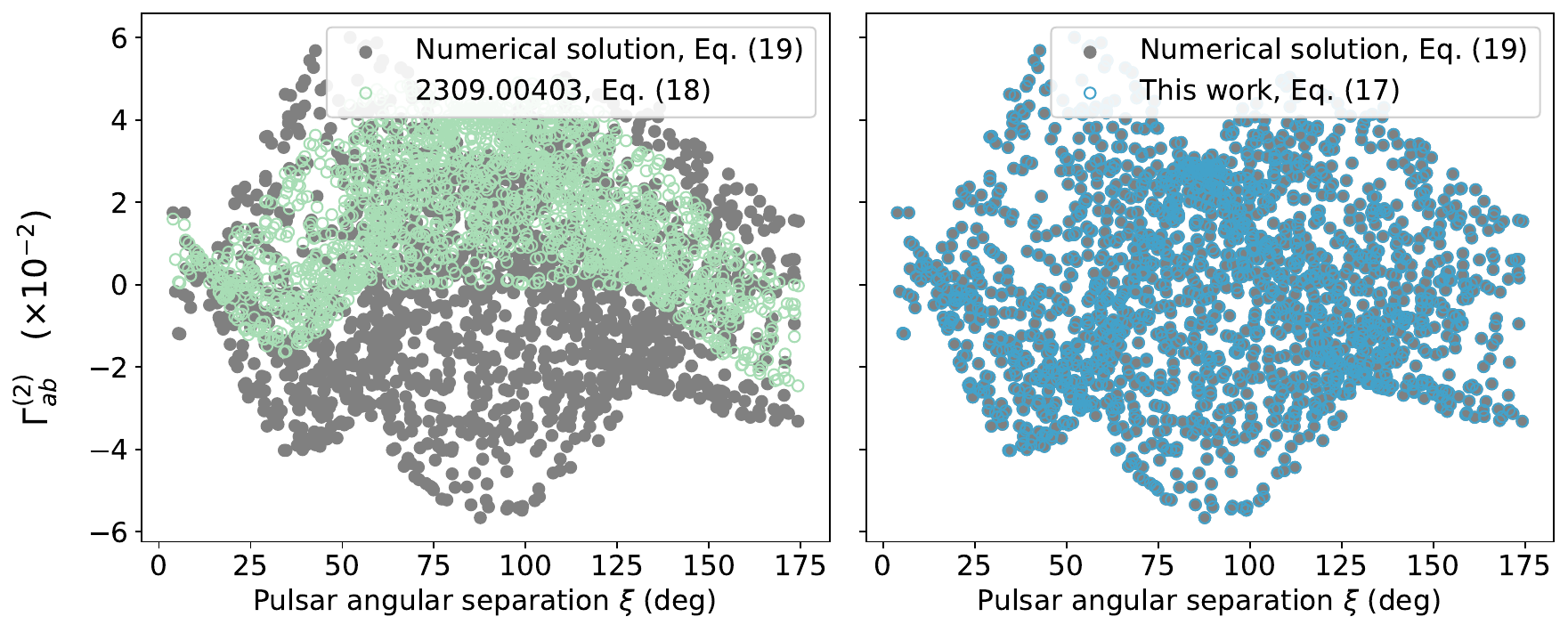}
    \caption{Comparison of the kinematic quadrupole contribution to the inter-pulsar cross-correlation as a function of the pulsar separation angle $\xi$. Left panel: Solution to \Eqref{eq:integralGamma2} computed via our numerical integration (grey), compared with that reported in Refs.~\cite{Tasinato:2023zcg,Cruz:2024svc} (teal). Right panel: Solution to \Eqref{eq:integralGamma2} computed via our numerical integration (grey), compared with our analytical solution (blue). The direction of the observer's motion has been set to the direction of the CMB dipole as inferred by Planck~\cite{Planck:2018nkj}. }
    \label{fig:ComparisonQuadrupole_General}
\end{figure}

To validate our analytical calculation of the kinematic quadrupole, we also numerically integrate \Eqref{eq:integralGamma2}. Specifically, we consider a set of 60 pulsars drawn from a uniform distribution across the sky, yielding 1770 distinct pairs. The direction of the observer's motion, $\hat{\boldsymbol{v}}$, is fixed to align with the direction of motion of the solar system, as inferred from the CMB dipole by Planck~\cite{Planck:2018nkj}. The numerical evaluation is performed by discretizing the sky using a {\tt HEALPix} grid with $N_{\rm side}$ = 32, corresponding to pixels with an area of approximately $10^{-3}$ square radians~\cite{Gorski_2005,Zonca:2019vzt}. For each {\tt HEALPix} pixel, we compute $\hat{\boldsymbol{n}}$ and the quantity $\sum_\lambda D_a^\lambda(\hat{\boldsymbol{n}}) D_b^\lambda(\hat{\boldsymbol{n}}) $. \cref{eq:integralGamma2} is then numerically evaluated by summing the integrand over all {\tt HEALPix} pixels and multiplying by the pixel area in square radians.

In Fig.~\ref{fig:ComparisonQuadrupole_General}, we compare the numerical result with our analytical expression in  \Eqref{eq:ourGamma2ResultimText} and
with the analytical expression of Refs.~\cite{Tasinato:2023zcg,Cruz:2024svc} in \Eqref{eq:tasinatoGamma2Result}.
As evident from the left panel of Fig.~\ref{fig:ComparisonQuadrupole_General}, the numerical evaluation (grey) of \Eqref{eq:integralGamma2} is inconsistent with the previously reported results (teal). In contrast, as shown in the right panel, our analytical expression (blue) for the kinematic quadrupole $\Gamma_{ab}^{(2)}$ shows excellent agreement with the numerical integration (grey), with a normalized mean-squared difference of $\sim 0.08\%$. The residual difference arises from the combined effects of truncation and rounding errors. 

\subsection{Auto-correlations}\label{sec:SecIIb}

So far, we have focused exclusively on cross-correlations. A straightforward approach to compute auto-correlations is to return to \Eqref{eq:fullORF3} and set $a=b$, which implies $x_{ab}=0$. The relevant angular integrals are the same as in the cross-correlation case, and keeping terms including $\mathcal{O}(\beta^2)$ yields
\begin{equation}
\begin{split}
\label{eq:auto-correlationBetaContributions}
\Gamma_{aa} 
&= \left[ 
        1 - \frac{\beta^2}{6}\left( 1 - n_I^2 - \alpha_I \right) 
    \right] 
    + \frac{1}{2}\beta (n_I - 1)\left( \hat{\boldsymbol{v}} \cdot \hat{\boldsymbol{x}}_a \right) \\
&\hspace{0.75in}
    + \frac{\beta^2}{2}\left( 2 - 3 n_I + n_I^2 + \alpha_I \right)
      \left(
        \frac{1}{10}\left( \hat{\boldsymbol{v}} \cdot \hat{\boldsymbol{x}}_a \right)^2 
        - \frac{1}{30}
      \right)\,.
\end{split}
\end{equation}
This approach, however, assumes $fL_a \to \infty$. Instead, we may keep the full dependence on $fL_a$:
\begin{align}\label{eq:fullAutoCorr}
   \Gamma_{aa} =  \frac{3}{4 \pi}\int d^2 \hat{\boldsymbol{n}} \, \sum_\lambda \left[D_a^\lambda (\hat{\boldsymbol{n}})\right]^2 \left[ 1- \cos\left(2\pi  f L_a (1+\hat{\boldsymbol{n}} \cdot\hat{\boldsymbol{x}}_a)\right) \right] \frac{I(f, \hat{\boldsymbol{n}})}{I'(f)} \, .
\end{align} 
Upon using the $\beta$-expansion of \Eqref{eq:taylorExpansionInBeta}, the angular integrals can be evaluated explicitly. For the monopole contribution, this computation was carried out in Ref.~\cite{Cordes:2024oem} and results in the expression:
\begin{tcolorbox}
\begin{equation}\label{eq:ourACZeroOrder}
    \Gamma_{aa}^{(0)} = 1 - \frac{3 \left[1- \mathrm{sinc} (4\pi fL_a) \right]}{8 (\pi fL_a)^2} \, .
\end{equation}
\end{tcolorbox}
To compute the higher multipole contributions to the auto-correlation, let us fix the pulsar direction along the $z$-axis: $\hat{\boldsymbol{x}}_a =\hat{\boldsymbol{z}}$. Moreover, let us choose the velocity direction $\hat{\boldsymbol{v}}$ to lie in the $xz$-plane and denote by $\alpha$ the angle between the pulsar direction and the velocity, so that
\begin{align}
\hat{\boldsymbol{v}}\cdot \hat{\boldsymbol{x}}_a &= \cos\alpha\,, &
    \hat{\boldsymbol{n}} \cdot\hat{\boldsymbol{x}}_a &= \cos \theta\,, & \hat{\boldsymbol{n}}\cdot \hat{\boldsymbol{v}} &= \cos \alpha \cos \theta + \sin \alpha \sin \theta \cos \phi \, .
\end{align}
With this choice, the sum containing the antenna pattern functions takes the simple form
\begin{equation}
    \sum_\lambda \left[D_a^\lambda(\hat{\boldsymbol{n}}) \right]^2 =  \frac{1}{4} \left( 1- \cos \theta \right)^2 \, .
\end{equation}
The dipole contribution to the auto-correlation function is then given by
\begin{tcolorbox}
\vspace{-0.5cm}
\begin{equation}\label{eq:ourACLinearOrder}
\begin{split}
    \Gamma_{aa}^{(1)} &= \frac{3}{16\pi} \int d^2 \hat{\boldsymbol{n}} \, \left[ 1-  \cos (2\pi fL_a (1+ \hat{\boldsymbol{n}}\cdot \hat{\boldsymbol{x}}_a))\right] \left( \hat{\boldsymbol{n}}\cdot \hat{\boldsymbol{v}} \right) \left( 1- \cos \theta \right)^2 \\
    &= \left[ - \frac{1}{2} + \frac{3}{4(\pi fL_a)^2} + \frac{3\sin(4 \pi fL_a)}{32(\pi fL_a)^3} + \frac{9(\cos(4 \pi fL_a) - 1)}{64(\pi fL_a)^4}\right]\left(\hat{\boldsymbol{v}} \cdot \hat{\boldsymbol{x}}_a \right) \, .
\end{split}
\end{equation}
\end{tcolorbox}
\noindent
The quadrupole contribution $\Gamma_{aa}^{(2)}$ can be computed in a similar way:
\begin{tcolorbox}
\vspace{-0.5cm}
\begin{equation}\label{eq:ourACQuadraticOrder}
\begin{split}
    \Gamma_{aa}^{(2)} &= \frac{3}{8 \pi} \int d^2 \hat{\boldsymbol{n}}\,
    \left[ 2- 2 \cos (2 \pi f L_a (1 + \hat{\boldsymbol{n}} \cdot\hat{\boldsymbol{x}}_a))\right] 
    \left[ (\hat{\boldsymbol{n}} \cdot\hat{\boldsymbol{v}})^2 - \frac{1}{3} \right] 
    \frac{1}{4} (1 - \cos \theta)^2  \\ 
    &= 
    \bigg[ \frac{16}{15} - \frac{16}{(\pi fL_a)^{2}} + 
    \frac{1}{(\pi fL_a)^{3}}\sin\left(4 \pi fL_a\right)  
    +\frac{9}{2(\pi fL_a)^{4}}\left[3+\cos(4\pi fL_a)\right]\\
    &\hspace{0.75in}
    - \frac{9}{2(\pi fL_a)^{5}}\sin\left(4 \pi fL_a\right)
    \bigg]
    \left(\frac{3 \left( \hat{\boldsymbol{v}} \cdot \hat{\boldsymbol{x}}_a \right)^2 - 1}{32} \right).
\end{split}
\end{equation}
\end{tcolorbox}
\noindent
As a consistency check, we note that in the limit $fL_a \to \infty$, the expressions in \Eqref{eq:ourACZeroOrder}, \Eqref{eq:ourACLinearOrder}, and \Eqref{eq:ourACQuadraticOrder} indeed correctly reproduce the result of \Eqref{eq:auto-correlationBetaContributions} that was directly computed in this limit. 

\begin{figure} 
    \isPreprints{\centering}{} 
    \includegraphics[width=0.75\linewidth]{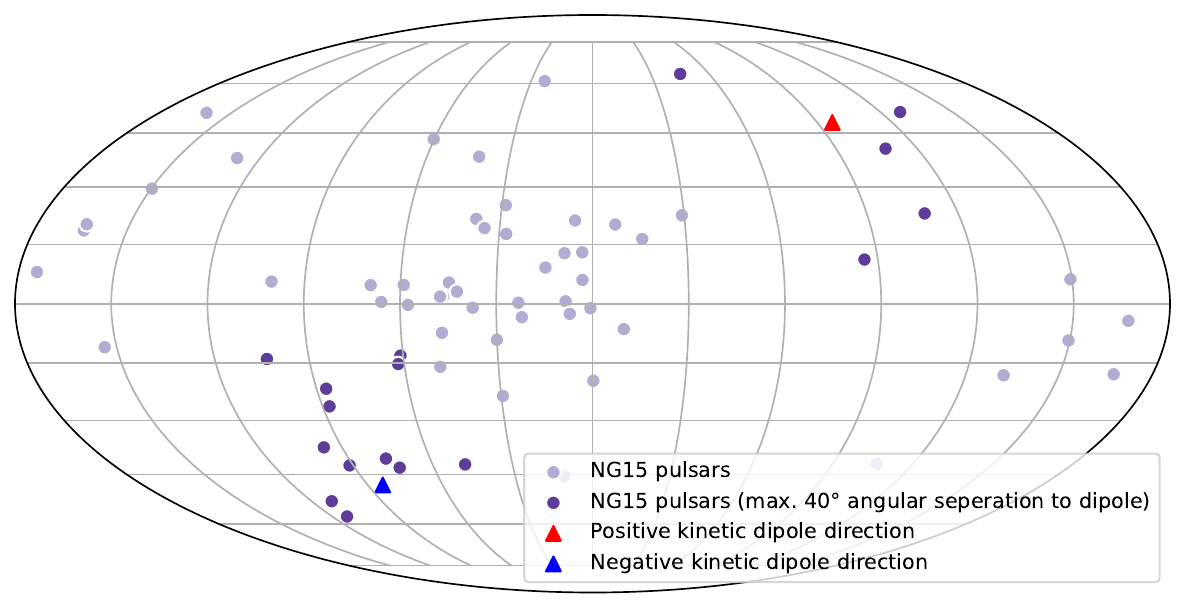}
    \caption{Location of the 67 pulsars that were used by the NANOGrav collaboration in~\cite{NANOGrav:2023gor}, shown here in galactic coordinates, and plotted using a Mollweide projection of the celestial sphere.    
    The direction of the velocity vector $\hat{\boldsymbol{v}}$ as determined by the CMB dipole and its antipodal point are depicted by a red triangle (galactic coordinates $(l,b) = (264^{\circ},48^{\circ})$ \cite{Planck:2018nkj}) and a blue triangle, respectively. Pulsars within an angular separation of $40^\circ$ from the direction $\pm\hat{\boldsymbol{v}}$ are shown in a darker tone. (Pulsar positions taken from Ref.~\cite{Manchester:2004bp}.)
    }
    \label{fig:pulsarMapping}
\end{figure}

\section{Effect of the kinematic dipole on the shape of the ORF}\label{sec:SecIIc}

We now illustrate the effect of our relative motion with respect to the isotropic-GWB frame on the ORF. For the expected value of $\beta = \mathcal{O}(10^{-3})$~\cite{Planck:2018nkj}, the quadrupole term will be strongly suppressed and hence negligible in phenomenological analyses that only allow for $\beta = \beta_{\rm CMB}$. We, however, stress that our results for the quadrupole contributions to the cross- and auto-correlations are, in fact, relevant for PTA data analyses that aim at setting an upper limit on $\beta$. Analyses of this type will also sample over $\beta$ values much larger than the CMB value, $\beta \gg \beta_{\rm CMB}$, which means that the determination of a reliable upper limit does depend on a proper understanding of the ORF up to higher orders in $\beta$. For instance, an upper limit close to $\beta \simeq 0.297$~\cite{Cruz:2024svc} requires control over the quadrupole contributions. 

\begin{figure}
    \isPreprints{\centering}{} 
    \includegraphics[width=0.75\linewidth]{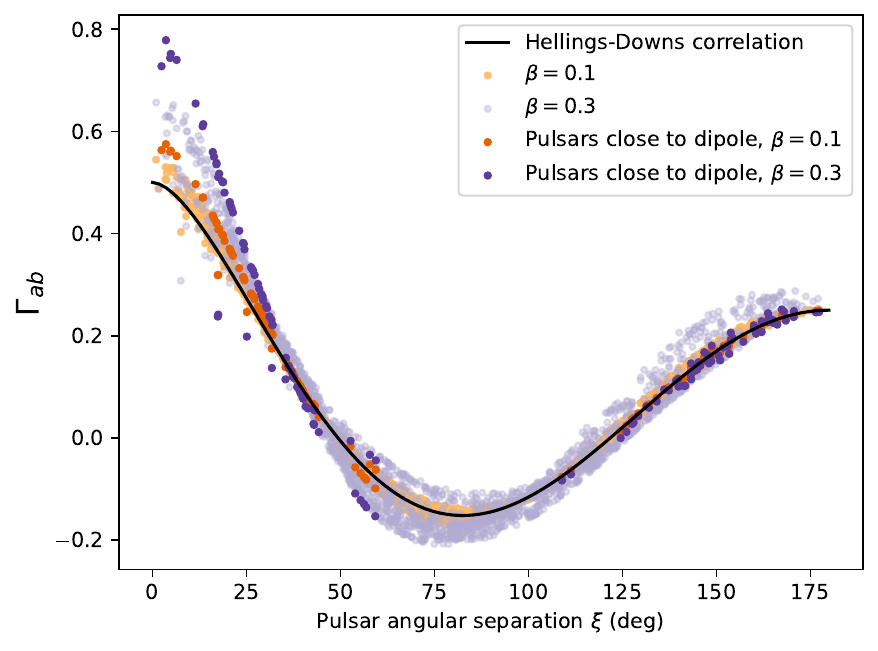}
    \caption{Plot showing the ORF $\Gamma_{ab}$ as a function of the pulsar-pair angular separation $\xi$, assuming a power-law GWB intensity with constant spectral index $n_I = -7/3$ ($\alpha_I=0$) as defined in \Eqref{eq:spectralParameter}. The direction $\hat{\boldsymbol{v}}$ is chosen to align with that inferred from the CMB dipole.
    For each pulsar pair formed from the 67 pulsars in Fig.~\ref{fig:pulsarMapping}, we show the corresponding value of the ORF as orange and purple dots for $\beta =0.1$ and $\beta = 0.3$, respectively. For pulsar pairs with both pulsars with an angular separation of at most $40^\circ$ from $\pm\hat{\boldsymbol{v}}$, we show the dots in dark orange and dark purple, respectively. For reference, we also plot the Hellings--Downs correlation ($\beta =0$) in black.}
    \label{fig:randomPulsar}
\end{figure}

We shall now consider the $67$ pulsars used in the NANOGrav 15-year dataset. Their positions on the sky are shown in Fig.~\ref{fig:pulsarMapping}. Each pulsar pair is associated with a fixed angular separation $\xi$. For all possible pulsar pairs obtained from these 67 pulsars, we compute $\Gamma_{ab}(\xi)$, assuming that the velocity aligns with the direction inferred from the CMB dipole, and consider the values $\beta = 0.1$ and $\beta =0.3$.\footnote{Recall that the the Taylor expansion in \Eqref{eq:taylorExpansionInBeta} is only valid for $\beta \ll 1$. In Fig.~\ref{fig:randomPulsar}, we work with $\beta = 0.1$ and $\beta =0.3$, for which the Taylor expansion is still (marginally) justified, in order to visually emphasize the respective trends in the ORF.} For each pulsar pair and velocity, we obtain a single value for $\Gamma_{ab}$, depicted as dots in Fig.~\ref{fig:randomPulsar}. For comparison, we also show the Hellings--Downs curve (black) given in Eq.~\eqref{eq:GammaHD}.

As expected, the overall shape of the ORF remains close to the Hellings–Downs curve, with larger deviations for $\beta = 0.3$ than for $\beta = 0.1$. The deviations can be significant even when the direction of one of the pulsars in a pair has a large angular separation from both $\hat{\boldsymbol{v}}$ and its antipodal direction. 

Note that for pulsar pairs with both pulsars located near the $\pm \hat{\boldsymbol{v}}$ direction, the largest deviations from the Hellings--Downs curve appear at small angles $\xi$ between the pulsars. In this case, both pulsars will either align with $\hat{\boldsymbol{v}}$, leading to larger values of the ORF at nonzero $\beta$, or with $-\hat{\boldsymbol{v}}$, leading to smaller values. This behavior follows directly from the dipole contribution in \Eqref{eq:dipoleHD}, which is linear in $\hat{\boldsymbol{v}}\cdot\hat{\boldsymbol{x}}_p$. In the former case, one obtains $\hat{\boldsymbol{v}} \cdot \hat{\boldsymbol{x}}_p\approx 1$ and thus a positive contribution to the ORF, whereas in the latter case yields $\hat{\boldsymbol{v}} \cdot \hat{\boldsymbol{x}}_p\approx -1$ and thus a negative deviation from the Hellings--Downs curve.
 
At large angular separations $\xi$ between the pulsars, we find the opposite effect. When both pulsars lie close to $\pm\hat{\boldsymbol{v}}$, one will satisfy $\hat{\boldsymbol{v}}\cdot\hat{\boldsymbol{x}}_a \approx 1$ while the other satisfies $\hat{\boldsymbol{v}}\cdot\hat{\boldsymbol{x}}_b \approx -1$. Consequently, the dipole contribution in \Eqref{eq:dipoleHD}, which depends on the sum of these two terms, becomes strongly suppressed. Small deviations from the Hellings–Downs curve nevertheless remain due to the quadrupole term, which is second order in $\hat{\boldsymbol{v}}\cdot\hat{\boldsymbol{x}}_p$ and therefore does not undergo the same cancellation.

\section{Conclusions}\label{sec:SecIII}

In this paper, we revisited the analytical treatment of PTA searches for kinematic anisotropies, building on the earlier work in Refs.~\cite{Cusin:2022cbb, Tasinato:2023zcg}. The central result of our analysis is that we found a quadrupole contribution to the cross-correlation ORF $\Gamma_{ab}^{(2)}$ that differs from the one reported in Refs.~\cite{Cusin:2022cbb, Tasinato:2023zcg}. For example, the result presented here does not vanish in the ``transverse-plane" geometry in which the pulsar positions lie in the plane orthogonal to the velocity vector $\hat{\boldsymbol{v}}$. We numerically evaluated the integral to show the validity of our result. 

Furthermore, we pointed out the relevance of the auto-correlation coefficient, which we discussed from two different perspectives: (i) The result in \Eqref{eq:auto-correlationBetaContributions} simply follows from the cross-correlation in \Eqref{eq:fullORF2} in the limit of identical pulsar positions, $x_{ab} \to 0$, and an additional factor of $1 + \delta_{ab} \to 2$. (ii) Meanwhile, the expressions in \Eqref{eq:ourACZeroOrder}, \Eqref{eq:ourACLinearOrder}, and \Eqref{eq:ourACQuadraticOrder} are derived from \Eqref{eq:fullAutoCorr}, which does not discard the oscillating pulsar terms in the integral. This second method is largely inspired by the treatment in Ref.~\cite{Cordes:2024oem}. Approaches (i) and (ii) agree with each other as expected in the limit $f L_a \to \infty$.

The presented toolkit can be used to set upper bounds on $\beta$ using PTA analysis tools such as PTArcade~\cite{Mitridate:2023oar}. It also allows for the construction of forecasts like those presented in Refs.~\cite{Tasinato:2023zcg,Cruz:2024svc}. 

\bigskip\noindent
{\small\textbf{Author Contributions:} Conceptualization, K.S.; Methodology, K.S.; Software, M.B.; Validation, D.A., J.D.R., T.S.; Formal Analysis, M.B.; Resources, K.S.; Writing – Original Draft Preparation, D.A., M.B., T.S.; Writing – Review \& Editing, J.D.R., K.S.; Visualization, D.A., M.B.; Supervision, K.S., T.S.; Project Administration, K.S.}

\medskip\noindent
{\small\textbf{Funding:} K.S.\ is an affiliate member of the Kavli Institute for the Physics and Mathematics of the Universe (Kavli IPMU) at the University of Tokyo and as such supported by the World Premier International Research Center Initiative (WPI), MEXT, Japan (Kavli IPMU). T.S.\ was funded by a CITA National Fellowship. D.A.\ and J.D.R.\ acknowledge financial support from NSF Physics Frontier Center Award PFC-2020265 and start-up funds from the University of Texas Rio Grande Valley.}

\medskip\noindent
{\small\textbf{Data Availability Statement:} All data generated during this project are visualized in the manuscript; see Figs.~\ref{fig:ComparisonQuadrupole}, \ref{fig:ComparisonQuadrupole_General}, and \ref{fig:randomPulsar}. The explicit numerical data behind these plots can be made available by the authors upon request.}

\medskip\noindent
{\small\textbf{Acknowledgments:} We thank Marisol Jiménez Cruz, Ameek Malhotra, Gianmassimo Tasinato, and Ivonne Zavala for friendly and fruitful discussions.}

\medskip\noindent
{\small\textbf{Conflicts of Interest:} The authors declare no conflict of interest. The sponsors had no role in the design, execution, interpretation, or writing of the study.}



\abbreviations{Abbreviations}{
The following abbreviations are used in this manuscript:

\noindent 
\begin{tabular}{@{}ll}
PTA & pulsar timing array\\
ORF & overlap reduction function\\
GW & gravitational wave\\
GWB & gravitational-wave background\\
CMB & cosmic microwave background
\end{tabular}
}

\appendixtitles{no} 
\appendixstart
\appendix

\section[\appendixname~\thesection]{Monopole Contribution} \label{appendix:monopole}
In this and the following two appendices, we present the full analytical calculation of the first three contributions to the overlap reduction function given in \Eqref{eq:Monopoledef} to \Eqref{eq:Quadrupoledef}.
We will start in this first appendix by obtaining the analytical expression of the Hellings-Downs curve, following the derivation given in Ref.~\cite{Jenet:2014bea}.

The integral we need to evaluate is
\begin{equation}
    \Gamma_{ab}^{(0)} = \frac{3}{8 \pi} \int d^2 \hat{\boldsymbol{n}} \, \sum_\lambda D_a^{\lambda}(\hat{\boldsymbol{n}}) D_b^{\lambda}(\hat{\boldsymbol{n}})\,.
\end{equation}
We will work in a coordinate system with pulsar $a$ located along the $z$-axis and pulsar $b$ in the $xz$-plane, where we identify the angle between the pulsars as $\xi$:
\begin{equation}
\hat{\boldsymbol{x}}_a \equiv \hat{\boldsymbol{z}}\,,
\qquad
\hat{\boldsymbol{x}}_b \equiv \sin\xi\,\hat{\boldsymbol{x}}+\cos\xi\,\hat{\boldsymbol{z}}\,.
\label{e:xa,xb}
\end{equation}
This leads to
\begin{align}   \hat{\boldsymbol{n}}\cdot\hat{\boldsymbol{x}}_a &= \cos \theta\,, \\    \hat{\boldsymbol{n}}\cdot\hat{\boldsymbol{x}}_b &= \cos \xi \cos \theta + \sin \xi \sin \theta \cos \phi\,,
\end{align}
where in this coordinate system $\hat{\boldsymbol{n}} = \hat{\boldsymbol{r}}$.
Recall that the polarization tensors $\mathbf{e}^\lambda_{ij} (\hat{\boldsymbol{n}})$ can be written in the following way:
\begin{equation}
\begin{aligned}
    e^+_{ij} (\hat{\boldsymbol{n}}) = \hat \theta_i \hat \theta_j - \hat \phi_i \hat \phi_j\,, \\
    e^\times_{ij} (\hat{\boldsymbol{n}}) =  \hat \theta_i \hat \phi_j + \hat \phi_i \hat \theta_j\,,
    \label{e:polarization_tensors}
\end{aligned}
\end{equation}
where $\hat{\boldsymbol{\phi}}$ and $\hat{\boldsymbol{\theta}}$ denote the unit vectors of the spherical coordinates. In order to evaluate the sum over the polarization index $\lambda$, we calculate the products of the pulsar direction vectors and the polarization tensors:
\begin{align}
    \hat x_a^{i} \hat x_a^j e^+_{ij} (\hat{\boldsymbol{n}}) &=  \sin^2 \theta\,, \\
    \hat x_a^{i} \hat x_a^j e^\times_{ij} (\hat{\boldsymbol{n}}) &= 0\,, \\
    \hat x_b^{i} \hat x_b^j e^+_{ij} (\hat{\boldsymbol{n}}) &=  \left( \sin \xi \cos \theta \cos \phi - \cos \xi \sin \theta \right)^2 - \sin^2 \xi \sin^2 \phi\,, \\
    \hat x_b^{i} \hat x_b^j e^\times_{ij} (\hat{\boldsymbol{n}}) &= 2 \left( \sin \xi \cos \theta \cos \phi - \cos \xi \sin \theta \right) \sin \xi \sin \phi\,.
\end{align}
It is clear that we can omit the $\times$-polarization in this coordinate system. It follows that:
\begin{align} \label{eq:daFromAppendix}
    D_a^+(\hat{\boldsymbol{n}}) &= \frac{1}{2} \left( 1- \cos \theta \right)\,, \\
    D_b^+(\hat{\boldsymbol{n}}) &= \frac{1}{2} \left[ (1 - \cos \xi \cos \theta - \sin \xi \sin \theta \cos \phi) - \frac{2 \sin^2 \xi \sin^2 \phi}{1 + \cos \xi \cos \theta + \sin \xi \sin \theta \cos \phi} \right]\,,
\end{align}
and
\begin{multline}
        \sum_\lambda D_a^\lambda(\hat{\boldsymbol{n}}) D_b^\lambda(\hat{\boldsymbol{n}}) =  \frac{1}{2} \left( 1- \cos \theta \right)
        \\\times
        \left[ \frac{1}{2} (1 - \cos \xi \cos \theta - \sin \xi \sin \theta \cos \phi)
        - \frac{\sin^2 \xi \sin^2 \phi}{1 + \cos \xi \cos \theta + \sin \xi \sin \theta \cos \phi} \right].
\end{multline}
All of the $\phi$-dependence is collected within the integral 
\begin{equation}
I(x ,\xi)\equiv \int_0^{2\pi}\,d\phi\>
\left[ \frac{1}{2} (1 - x\cos \xi - \sqrt{1-x^2}\sin \xi\cos \phi)
        - \frac{\sin^2 \xi \sin^2 \phi}{1 + x\cos \xi + \sqrt{1-x^2}\sin \xi\cos \phi} \right]\,,
        \label{e:Iint}
\end{equation}
where we substituted $x \equiv \cos \theta$ under the integral, leaving us with
\begin{equation}
    \Gamma_{ab}^{(0)} = \frac{3}{16 \pi} \int_{-1}^{1} dx \, (1-x) I(x, \xi)
\end{equation}
to carry out.
In order to evaluate $I(x, \xi)$, we split it into two parts
\begin{equation}
    I(x, \xi) =I_1(x, \xi) + I_2(x,\xi)\,,
    \end{equation}
of which one,
\begin{equation}
    I_1 (x,\xi) \equiv \frac{1}{2} \int_0^{2\pi} d \phi \, (1- x \cos \xi - \sqrt{1 - x^2} \sin \xi \cos \phi) = \pi (1 - x \cos \phi)\,,
\end{equation}
is trivially solved, and the second,
\begin{equation}
     I_2 (x, \xi) \equiv - \sin^2 \xi \int_0^{2\pi} d \phi \, \frac{\sin^2 \phi}{1 + x \cos \xi + \sqrt{1 - x^2} \sin \xi \cos \phi} \\
     = - \sin^2 \xi \oint_{C} dz \, f(z)
\end{equation}
can be solved using contour integration.
When writing the contour integral for $I_2(x,\xi)$, we make the substitution $z = \exp (i \phi)$. This leaves us with the following:
\begin{equation}
    f(z) = \frac{i (z^2 - 1)^2 }{z^2 \left[ 4 z (1 + x \cos \xi) + 2 \sqrt{1 - x^2} \sin \xi (z^2 +1) \right]}\,.
\end{equation}
The part of the denominator of $f(z)$ that is enclosed in square brackets can be written in terms of its poles. This can be done by factoring out $2 \sqrt{1 - x^2} \sin \xi$ and applying the quadratic equation in terms of $z$:
\begin{equation}
    4 z (1 + x \cos \xi) + 2 \sqrt{1 - x^2} \sin \xi (z^2 + 1) = 2 \sqrt{1 - x^2} \sin \xi (z - z_+) (z - z_+)\,.
\end{equation}
When applying the quadratic formula, the term $|x+\cos \xi|$ arises; therefore, we divide the integral over $x$ into two parts for two regions. In the following, the integral from $- \cos \xi \leq x \leq 1$ (corresponding to $x+\cos\xi\ge 0$) will refer to the top sign, and $-1 \leq x \leq - \cos \xi$ (corresponding to $x+\cos\xi\le 0$) to the bottom sign:
\begin{equation}\label{eq_defZ+Z-}
    z_+ \equiv -\sqrt{\left( \frac{1 \mp \cos \xi}{1 \pm \cos \xi} \right) \left( \frac{1 \mp x}{1 \pm x} \right)}\,,\qquad z_- \equiv \frac{1}{z_+}\,.
\end{equation}
To see which of these poles is within the unit circle to calculate the residue, we look at their absolute value and estimate from there:
\begin{equation}
    |z_+| = \left| \sqrt{\left( \frac{1 \mp \cos \xi}{1 \pm \cos \xi} \right) \left( \frac{1 \mp x}{1 \pm x} \right)} \right| = \left| \sqrt{\frac{1 + x \cos \xi \mp(x+\cos\xi)}{1 + x\cos \xi \pm (x+\cos\xi)}} \right| \leq 1\,,
\end{equation}
which is valid in both regions $x+\cos\xi\ge 0$ (top sign) and $x+\cos\xi\le 0$ (bottom sign).
Thus, $z_+$ lies within the unit circle, while $|z_-|=1/|z_+|$ lies outside.

To summarize, until this point we have found two poles of $f(z)$, namely $z_+$ and $0$, the latter of which is a pole of order two. The definition of a residue of order $n$ at $z_0$ is the following:
\begin{equation}
    \mathrm{Res}(f(z), z_0) = \lim_{z \to z_0} \frac{1}{(n - 1)!} \frac{d^{n-1}}{dz^{n-1}} (f(z) (z-z_0)^n)\,,
\end{equation}
leaving us with
\begin{equation}
    \mathrm{Res}(f(z), z_+) = \frac{i (z_+ - z_-)}{2 \sqrt{1- x^2}\,\sin \xi}\,, \qquad
    \mathrm{Res}(f(z), 0) = \frac{i (z_+ + z_-)}{2 \sqrt{1-x^2}\,\sin \xi}\,.
\end{equation}
Applying the residue theorem gives the result for $I_2$:
\begin{equation}
    \oint_{C} f(z) dz = 2 \pi i \sum_{i} \mathrm{Res}(f, z_i) = \frac{2 \pi}{(1 \pm x)(1 \pm \cos \xi)}\,,
\end{equation}
and putting everything together:
\begin{equation}\label{eq_INoDipol}
    I(x, \xi) = \pi (1- x \cos \xi) - 2 \pi \frac{1 \mp \cos \xi}{1 \pm x}\,.
\end{equation}
In the last step, we carry out the integration over $x$ where we consider our defined regions, therefore our integral splits into three parts:
\begin{equation}
\begin{split}
    \Gamma_{ab}^{(0)} &= \frac{3}{16} \int_{-1}^{1} dx \, (1-x) (1-x \cos \xi) 
    - \frac{3}{8} \left[ (1 + \cos \xi) \int_{-1}^{-\cos \xi} dx + (1-\cos \xi) \int_{- \cos \xi}^{1} dx \, \frac{1-x}{1+x} \right] \\
    &= \frac{1}{2} - \frac{1}{4} \left( \frac{1 - \cos \xi}{2} \right) + \frac{3}{2} \left( \frac{1 - \cos \xi}{2} \right) \ln \left( \frac{1 - \cos \xi}{2} \right) \\
    &= \frac{1}{2} - \frac{1}{4} x_{ab} + \frac{3}{2} x_{ab} \ln \left( x_{ab} \right)\,,
\end{split}
\end{equation}
where we have defined
\begin{equation}
    x_{ab} \equiv \frac{1 - \cos \xi}{2} \equiv \frac{1 - \hat{\boldsymbol{x}}_a \cdot\hat{\boldsymbol{x}}_b}{2}\,.
\end{equation}

\section[\appendixname~\thesection]{Dipole Contribution}
\label{appendix:dipole}

The dipole contribution enters the integral with a factor of $\hat{\boldsymbol{n}} \cdot\hat{\boldsymbol{v}}$, where $\hat{\boldsymbol{v}}$ is the unit vector of the velocity of our Barycenter's moving reference frame (however, in theory it can be any vector for the computation). The derivation follows the example in Ref.~\cite{Anholm:2008wy} but uses methods similar to those used in the previous section.

The relevant integral looks like this:
\begin{equation}
    \Gamma_{ab}^{(1)} = \frac{3}{8 \pi} \int d^2 \hat{\boldsymbol{n}} \,  (\hat{\boldsymbol{n}} \cdot\hat{\boldsymbol{v}})\sum_\lambda D_a^{\lambda}(\hat{\boldsymbol{n}})D_b^{\lambda}(\hat{\boldsymbol{n}})\,,
\end{equation}
where we define the general direction of $\hat{\boldsymbol{v}}$ to be
\begin{equation} \label{eq:vinspherical}
    \hat{\boldsymbol{v}} \equiv (\sin \alpha_1 \cos \eta, \sin \alpha_1 \sin \eta, \cos \alpha_1)\,.
\end{equation}
Thus,
\begin{align} 
\hat{\boldsymbol{n}}\cdot\hat{\boldsymbol{v}} &= \cos \alpha_1 \cos \theta + \sin \alpha_1 \sin \theta \cos (\phi - \eta)\,,\label{eq:nvDefiniton}\\
\hat{\boldsymbol{v}}\cdot\hat{\boldsymbol{x}}_a &= \cos \alpha_1 \,,\label{eq_cosalpha_1}\\
\hat{\boldsymbol{v}}\cdot\hat{\boldsymbol{x}}_b &= \cos \alpha_1 \cos \xi + \sin \alpha_1 \sin \xi \cos \eta \equiv \cos \alpha_2\,.\label{eq_cosalpha_2}
\end{align}
Plugging everything into the integral, we obtain the following:
\begin{multline}
    \Gamma_{ab}^{(1)} = \frac{3}{16 \pi} \int d^2 \hat{\boldsymbol{n}} \, \left[ \cos \alpha_1 \cos \theta + \sin \alpha_1 \sin \theta (\cos \eta \cos \phi + \sin \phi \sin \eta) \right] \\
    \times\frac{1}{2} (1- \cos \theta) \left[ (1- \cos \xi \cos \theta  - \sin \xi \sin \theta \cos \phi ) - \frac{2 \sin^2 \xi \sin^2 \phi}{1+ \cos \xi \cos \theta + \sin \xi \sin \theta \cos \phi} \right]\,.
\end{multline}
We once again make the substitution $x = \cos \theta$ and observe that the integral can be split into three parts. Each part is defined by the prefactor within the first square brackets:
\begin{multline}
    \Gamma_{ab}^{(1)} = \frac{3}{16 \pi} \int_{-1}^{1} dx \, (1 - x) \\\times\left[ \cos \alpha_1 x \,I (x, \xi) + \sin \alpha_1 \sqrt{1-x^2} \cos \eta\, J(x, \xi) + \sin \alpha_1 \sqrt{1-x^2} \sin \eta\, \tilde J(x, \xi) \right]\,,
\end{multline}
where $I(x,\xi)$ is the same as in \Eqref{e:Iint} and \Eqref{eq_INoDipol},
and $J(x,\xi)$ and $\tilde J(x,\xi)$ are the same integral except for a factor of $\cos \phi$ and $\sin \phi$, respectively (see below). Each of these integrals can be further split apart following the example in the previous section ($J \equiv J_{1} + J_{2}$, $\tilde J \equiv \tilde J_{1} + \tilde J_{2}$):
\begin{align}
    J_{1} &\equiv \frac{1}{2} \int_0^{2 \pi} d \phi \, \cos \phi (1- x \cos \xi - \sqrt{1 -x^2} \sin \xi \cos \phi) = -\frac{\pi}{2} \sqrt{1 -x^2} \sin \xi\,, \\
    J_{2} &\equiv - \sin^2 \xi \int_0^{2 \pi} d \phi \, \frac{\sin^2 \phi \cos \phi}{1 + x \cos \xi + \sqrt{1-x^2} \sin \xi \cos \phi}\,, \\
    \tilde J_{1} &\equiv \frac{1}{2} \int_0^{2 \pi} d \phi \, \sin \phi (1- x \cos \xi - \sqrt{1 -x^2} \sin \xi \cos \phi) = 0\,, \\
    \tilde J_{2} &\equiv - \sin^2 \xi \int_0^{2 \pi} d \phi \, \frac{\sin^3 \phi}{1 + x \cos \xi + \sqrt{1-x^2} \sin \xi \cos \phi}=0\,,
\end{align}
where the $\tilde J_1$ and $\tilde J_2$ integrals vanish because periodicity of the integrands allow us to change the limits from $0$ to $2\pi$ to $-\pi$ to $\pi$ (even interval), but the integrands are odd functions of $\phi$.
Thus, $\tilde J=0$.

To evaluate the $J_2$ integral, we once again make use of the substitution $z = \exp (i \phi)$ similar to the previous subsection and use the same definitions as in \Eqref{eq_defZ+Z-}:
\begin{align}
    J_{2} &= - \frac{\sin^2 \xi}{2} \oint_C dz \, \underbrace{\frac{i(z^2-1)^2 (z^2+1)}{ 2 \sqrt{1-x^2}\, \sin \xi\, z^3 (z - z_+)(z - z_-)}}_{f_{2}(z)}\,.
\end{align}
Observing that the residue at pole $z_0 = 0$ is now of order three, we find
\begin{equation}
    \mathrm{Res}(f_{2}(z), 0) = \frac{i(z_+^2 + z_-^2)}{2 \sqrt{1-x^2}\,\sin \xi}\,,\qquad
    \mathrm{Res}(f_{2}(z), z_+) = \frac{i (z_+^2 - z_-^2)}{2 \sqrt{1-x^2}\,\sin \xi}\,,\\
\end{equation}
and thus
\begin{align}
    J_{2} &= - \frac{\sin^2 \xi}{2} \,2 \pi i \sum_i \mathrm{Res}(f_{2}(z), z_i) = z_+^2 \frac{\pi \sin \xi}{\sqrt{1-x^2}} \,,
\end{align}
resulting in
\begin{align} \label{eq:J2Dipole}
    J(x,\xi) &= -\frac{\pi}{2} \sqrt{1 -x^2} \sin \xi + \left(\frac{1 \mp \cos \xi}{1 \pm \cos \xi}\right) \left(\frac{1\mp x}{1 \pm x}\right) \frac{\pi \sin \xi}{\sqrt{1-x^2}}\,.
\end{align}
Finally, we can put everything together and once again integrate over $x$, where we respect the established regions of $-1\le x\le - \cos \xi$ and $- \cos \xi\le x\le 1$:
\begin{equation}
\begin{split}
    \Gamma_{ab}^{(1)} &= \frac{3}{16}\bigg\{ \int_{-1}^{1} dx \, (1-x) \bigg[\cos \alpha_1 x \left((1-x \cos \xi) - 2 \frac{1 \mp \cos \xi}{1 \pm x}\right) \\
    &\hspace{0.5in}+ \sin \alpha_1 \sqrt{1-x^2} \cos \eta  \bigg( -\frac{1}{2} \sqrt{1-x^2} \sin \xi 
    + \left( \frac{1 \mp \cos \xi}{1 \pm \cos \xi} \right) \left( \frac{1 \mp x}{1 \pm x} \right)  \frac{\sin \xi}{\sqrt{1 -x^2}} \bigg) \bigg] \bigg\} \\
    &= \frac{3}{16} \bigg\{ \cos\alpha_1\bigg[ \int_{-1}^1 dx \, (1-x) x (1-x \cos \xi) \\
    &\hspace{0.5in}- 2 \int_{-1}^{-\cos \xi} dx \, x (1+ \cos \xi)  -2 \int_{-\cos \xi}^{1} dx \, x (1 - \cos \xi) \frac{1-x}{1+x}\bigg] \\
    &\hspace{0.5in}+\sin \alpha_1 \cos \eta \sin \xi  \bigg[ -\frac{1}{2} \int_{-1}^{1} dx \, (1-x) (1-x^2) \\
    &\hspace{0.5in}+ \int_{-1}^{-\cos \xi} dx \, \left( \frac{1+ \cos \xi}{1 - \cos \xi} \right) (1+x) + \int_{-\cos \xi}^{1} dx \, \left( \frac{1- \cos \xi}{1 + \cos \xi} \right) \frac{(1-x)^2}{1+x} \bigg]\bigg\} \\
    &= \frac{3}{16} \bigg\{ \cos \alpha_1 (\cos \xi + 1) \bigg[ - \frac{2}{3} + (1-\cos^2 \xi) \\
    &\hspace{0.5in}- (1- \cos \xi) (\cos \xi +3)-4 \frac{(1- \cos \xi)}{(1+\cos\xi)} \ln \left({\frac{1 - \cos \xi}{2}}\right)  \bigg] \\
    &\hspace{0.5in}+\sin \alpha_1 \cos \eta  \sin \xi \bigg[ -\frac{2}{3} - 2 (1- \cos \xi) - 4 \frac{1- \cos \xi}{1+ \cos \xi} \ln \left( \frac{1 - \cos \xi}{2} \right) \bigg]\bigg\} \\
    &=\frac{3}{16} \bigg\{ \left( \cos \alpha_1 (\cos \xi + 1) + \sin \alpha_1 \cos \eta \sin \xi\right) \\
    &\hspace{0.5in}\times\left[- \frac{2}{3} - 2 (1- \cos \xi) - 4 \frac{1 - \cos \xi}{1 + \cos \xi} \ln \left( \frac{1 - \cos \xi}{2}\right)\right]\bigg\} \\
    &= \left[\hat{\boldsymbol{v}}\cdot \hat{\boldsymbol{x}}_a + \hat{\boldsymbol{v}}\cdot \hat{\boldsymbol{x}}_b \right] \left( -\frac{1}{8} - \frac{3}{4}x_{ab} - \frac{3}{4}\frac{ x_{ab} }{\left( 1- x_{ab}\right)}\ln x_{ab}\right),
\end{split}
\end{equation}
where in the last step, we applied \Eqref{eq_cosalpha_1} and \Eqref{eq_cosalpha_2}.

\section[\appendixname~\thesection]{Quadrupole Contribution}
 \label{appendix:quadrupole}

The quadrupole contribution requires us to evaluate the following integral:
\begin{equation}
    \Gamma_{ab}^{(2)} = \frac{3}{8 \pi} \int d^2 \hat{\boldsymbol{n}} \, \left[ \left( \hat{\boldsymbol{n}}\cdot \hat{\boldsymbol{v}} \right)^2 - \frac{1}{3} \right] \sum_\lambda D_a^\lambda(\hat{\boldsymbol{n}}) D_b^\lambda(\hat{\boldsymbol{n}})\,.
\end{equation}
Using the same definitions as in \Eqref{eq:nvDefiniton}, one arrives at
\begin{equation}
\begin{split}
    \Gamma_{ab}^{(2)} &= \frac{3}{16 \pi} \int d^2 \hat{\boldsymbol{n}} \bigg[ \cos^2 \alpha_1 \cos^2 \theta + \sin^2 \alpha_1 \sin^2 \theta \left( \cos \phi \cos \eta + \sin \phi \sin \eta \right)^2 \\
    &\hspace{0.5in}+ 2 \cos \alpha_1 \sin \alpha_1 \sin \theta \cos \theta \left(\cos \phi \cos \eta + \sin \phi \sin \eta\right) - \frac{1}{3}\bigg]  \\
    &\hspace{0.5in}\times\frac{1}{2} (1 - \cos \theta)\left[ (1- \cos \xi \cos \theta  - \sin \xi \sin \theta \cos \phi ) - \frac{2 \sin^2 \xi \sin^2 \phi}{1+ \cos \xi \cos \theta + \sin \xi \sin \theta \cos \phi} \right] \\
    &= \frac{3}{16 \pi} \int_{-1}^{1} dx \, (1 - x) \bigg\{ \left[ \cos^2 \alpha_1 x^2 + \sin^2 \alpha_1 \cos^2 \eta (1- x^2) - \frac{1}{3} \right] I(x, \xi) \\
    &\hspace{0.5in}+ \sin^2 \alpha_1 (1-x^2) \left[ (\sin^2 \eta - \cos^2 \eta) K(x, \xi) + 2 \cos \eta \sin \eta\, \tilde{K}(x, \xi) \right] \\
    &\hspace{0.5in}+ 2 \cos \alpha_1 \sin \alpha_1\, x \sqrt{1- x^2} \left[ \cos \eta \,J(x, \xi) + \sin \eta \,\tilde J (x, \xi)  \right] \bigg\}\,,
\end{split}
\end{equation}
where after the second equal sign, we replaced $\cos \theta$ by $x$ and collected all $\phi$-dependencies in $I(x, \xi)$, $J(x, \xi)$, $\tilde J(x,\xi)$, $K(x, \xi)$ and $\tilde{K}(x, \xi)$. Once again $I(x, \xi)$ is \Eqref{eq_INoDipol}, while $J(x, \xi)$ is given in \Eqref{eq:J2Dipole} and $\tilde J(x, \xi)=0$.
Also note that $\tilde{K}(x, \xi)=0$, which follows by making the same argument that showed that $\tilde J(x,\xi)=0$. 

To evaluate $K(x,\xi)$ we split it into two parts, $K(x, \xi) = K_1 (x, \xi) + K_2 (x, \xi)$, and perform similar integrations as above:
\begin{align}
    K_1 &\equiv \int_0^{2 \pi} d \phi \, \frac{1}{2} \sin^2 \phi \left( 1 - x \cos \xi - \sqrt{1 - x^2} \sin \xi \cos \phi \right) = \frac{\pi}{2} (1 - x \cos \xi)\,, \\
    K_2 &\equiv - \sin^2 \xi \int_0^{2 \pi} d \phi \, \frac{\sin^4 \phi}{1 + x \cos \xi + \sin \xi \sqrt{1 - x^2} \cos \phi} = -\sin^2\xi\oint_{C} f(z) dz\,.
\end{align}
For $K_2$, the function that appears in the contour integral is
\begin{equation}
    f(z) = \frac{1}{4 i} \frac{\left(z^2 - 1 \right)^4}{2 \sqrt{1 - x^2} \sin \xi \left( z - z_+ \right) \left( z - z_- \right) z^4}\,,
\end{equation}
with the residues
\begin{equation}
    \mathrm{Res}(f(z), 0) = \frac{1}{8 i} \frac{z_+^3 - 3z_+ - 3 z_- + z_-^3}{\sqrt{1 - x^2} \sin \xi} \,,\qquad
    \mathrm{Res}(f(z), z_+) = \frac{1}{8 i} \frac{(z_+ - z_-)^3}{\sqrt{1 - x^2} \sin \xi}\,,
\end{equation}
such that $K(x, \xi)$ evaluates to
\begin{equation}
    K(x, \xi) = \frac{\pi}{2} (1- x \cos \xi) + \\
    \frac{\pi \sin \xi}{2 \sqrt{1 - x^2}} \sqrt{\left(\frac{1 \mp \cos \xi}{1 \pm \cos \xi}\right) \left( \frac{1 \mp x}{1 \pm x} \right)} \left[ \left(\frac{1 \mp \cos \xi}{1 \pm \cos \xi}\right) \left( \frac{1 \mp x}{1 \pm x} \right) - 3 \right].
\end{equation}
Our integral must still be evaluated over $x$, where we respect the dedicated regions and drag the respective terms with $x$ dependencies into the integral
\begin{equation}
\begin{split}
    &\frac{3}{16} \int_{-1}^{1} dx \, (1-x)\bigg\{\left[x^2 \cos^2 \alpha_1 + (1- x^2) \sin^2 \alpha_1 \cos^2 \eta - \frac{1}{3} \right] \left(1- x \cos \xi - 2 \frac{1 \mp \cos \xi}{1 \pm x} \right) \\
    &\hspace{0.5in}+ \sin^2 \alpha_1 \left[\sin^2 \eta - \cos^2 \eta \right] \bigg( \frac{1}{2} (1 - x \cos \xi) (1-x^2) \\
    &\hspace{1in}+ \frac{1}{2} (1 \mp \cos \xi) (1 \mp x) 
    \left[\left(\frac{1 \mp \cos \xi}{1 \pm \cos \xi}\right) \left(\frac{1 \mp x}{1 \pm x}\right) - 3\right]\bigg) \\
    &\hspace{0.5in}+ 2 \sin \alpha_1 \cos \alpha_1 \sin \xi \cos \eta \left( - \frac{1}{2} x (1 -x^2) + x\left(\frac{1 \mp \cos \xi}{1 \pm \cos \xi}\right) \left(\frac{1 \mp x}{1 \pm x}\right) \right)\bigg\} \\
    &= \frac{3}{16} \bigg\{\left( 8 x_{ab} \ln x_{ab} - \frac{2}{3} \cos^3 \xi -2 \cos^2 \xi + \frac{64}{60} \cos \xi + \frac{8}{3}\right) \cos^2 \alpha_1 \\
    &\hspace{0.5in}+ \left(-\frac{2}{3} - \frac{2}{5} \cos \xi + 2 \cos^2 \xi + \frac{2}{3} \cos^3 \xi \right) \sin^2 \alpha_1 \cos^2 \eta \\
    &\hspace{0.5in}- \left(8 x_{ab} \ln x_{ab} + \frac{2}{3} \cos \xi + 2\right) \frac{1}{3} \\
    &\hspace{0.5in}+ \sin^2 \alpha_1  \left[\sin^2 \eta - \cos^2 \eta\right] \bigg\{\left(\frac{2}{15} \cos \xi + \frac{2}{3}\right) \\
    &\hspace{0.5in}+\frac{4}{\cos \xi + 1}\left(-(1-\cos \xi)^2 \ln x_{ab} + \cos^2 \xi - 1 \right)\bigg\} \\
    &\hspace{0.5in}+ 2 \sin \alpha_1 \cos \alpha_1 \sin \xi \cos \eta \left(\frac{2}{15} + \frac{2}{3(1 + \cos \xi)} \left(12 x_{ab} \ln x_{ab} + \sin^2 \xi \cos \xi + 4\sin^2 \xi \right)  \right)\bigg\}\,.
    \label{e:A:62}
\end{split}
\end{equation}
To proceed further, we use the definitions of $\hat{\boldsymbol{v}}\cdot \hat{\boldsymbol{x}}_a$ and $\hat{\boldsymbol{v}}\cdot\hat{\boldsymbol{x}}_b$ to re-express the $\alpha_1$ and $\eta$-dependent factors that appear in the final line of the above expression:
\begin{align}
\cos^2\alpha_1 &= (\hat{\boldsymbol{v}}\cdot\hat{\boldsymbol{x}}_a)^2\,,
\label{e:cos2alpha}
\\
\sin^2\alpha_1\cos^2\eta &= \frac{1}{4x_{ab}(1-x_{ab})}\left[
(\hat{\boldsymbol{v}}\cdot\hat{\boldsymbol{x}}_a)^2\cos^2\xi + (\hat{\boldsymbol{v}}\cdot\hat{\boldsymbol{x}}_b)^2-
2(\hat{\boldsymbol{v}}\cdot\hat{\boldsymbol{x}}_a)(\hat{\boldsymbol{v}}\cdot\hat{\boldsymbol{x}}_b)\cos\xi
\right]\,,
\label{e:sin2alpha_cos2eta}\\
\sin^2\alpha_1\left[\sin^2\eta-\cos^2\eta\right] &= 1- (\hat{\boldsymbol{v}}\cdot\hat{\boldsymbol{x}}_a)^2-2\sin^2\alpha_1\cos^2\eta\,,
\label{e:sin2[sin2-cos2]}\\
\sin\alpha_1\cos\alpha_1\sin\xi\cos\eta &= 
(\hat{\boldsymbol{v}}\cdot\hat{\boldsymbol{x}}_a)\cdot(\hat{\boldsymbol{v}}\cdot\hat{\boldsymbol{x}}_b)-(\hat{\boldsymbol{v}}\cdot\hat{\boldsymbol{x}}_a)^2\,\cos\xi\,,
\label{e:sin_cos_sin_cos}
\end{align}
where we use \Eqref{e:sin2alpha_cos2eta} to expand the $2\sin^2\alpha_1\cos^2\eta$ term on the rhs of \Eqref{e:sin2[sin2-cos2]}.
Substituting these expressions into \Eqref{e:A:62}, and combining all terms multiplying $(\hat{\boldsymbol{v}}\cdot\hat{\boldsymbol{x}}_a)^2$,
$(\hat{\boldsymbol{v}}\cdot\hat{\boldsymbol{x}}_b)^2$,
and
$(\hat{\boldsymbol{v}}\cdot\hat{\boldsymbol{x}}_a)(\hat{\boldsymbol{v}}\cdot\hat{\boldsymbol{x}}_b)$, leads to the final expression:
\begin{equation}\label{eq:ourGamma2Result}
\begin{split}
\Gamma_{ab}^{(2)} 
&= \frac{1}{\left( x_{ab} -1 \right)^2} \bigg\{
    \left( 
        \frac{1}{4}x_{ab}^3 - x_{ab}^2 + \frac{3}{4} x_{ab} \ln x_{ab} 
        + \frac{1}{2} x_{ab} + \frac{1}{4} 
    \right) 
    \left[ 
        (\hat{\boldsymbol{v}}\cdot\hat{\boldsymbol{x}}_a)^2 
        + (\hat{\boldsymbol{v}}\cdot\hat{\boldsymbol{x}}_b)^2
    \right] \\
&\hspace{0.5in} 
    + \left( 
        \frac{3}{2} x_{ab}^2 \ln x_{ab} - \frac{39}{20} x_{ab}^2 
        + \frac{12}{5} x_{ab} - \frac{9}{20}
    \right) 
    \left[
        (\hat{\boldsymbol{v}}\cdot\hat{\boldsymbol{x}}_a)
        (\hat{\boldsymbol{v}}\cdot\hat{\boldsymbol{x}}_b)
    \right] \\
&\hspace{0.5in}
    + \left(
        x_{ab}^3 \ln x_{ab} - \frac{22}{15} x_{ab}^3 
        - \frac{1}{2} x_{ab}^2 \ln x_{ab} + \frac{35}{12} x_{ab}^2 
        - \frac{1}{2} x_{ab} \ln x_{ab} - \frac{43}{30} x_{ab} 
        - \frac{1}{60}
    \right)
\bigg\}\,.
\end{split}
\end{equation}

\isPreprints{}{
\begin{adjustwidth}{-\extralength}{0cm}
} 

\reftitle{References}


\bibliography{arxiv_v2.bib}

%


\isPreprints{}{
\end{adjustwidth}
} 
\end{document}